\documentclass[a4paper,11pt]{article}
\pdfoutput=1 

\usepackage{docstyle} 

\usepackage[T1]{fontenc} 
\usepackage{amsmath, amssymb, amsthm}
\usepackage{physics}
\usepackage{booktabs}

\usepackage{enumitem}
\usepackage{slashed}
\usepackage{revsymb}
\usepackage{stmaryrd}

\definecolor{mycolor}{RGB}{102, 0, 255}
\usepackage[colorlinks=true
,urlcolor=mycolor
,anchorcolor=mycolor
,citecolor=mycolor
,filecolor=mycolor
,linkcolor=mycolor
,menucolor=mycolor
,pagecolor=mycolor
,linktocpage=true
,pdfproducer=medialab
,pdfa=true
]{hyperref}

\usepackage{lmodern}

\newcommand{\pmuu}{\partial^\mu}
\newcommand{\pmud}{\partial_\mu}
\newcommand{\mc}[1]{\mathcal{#1}}
\newcommand{\lagr}{\mathcal{L}}
\newcommand{\half}{\frac{1}{2}}

\title{\textcolor{mycolor}{\boldmath Fate of Vainshtein Screening for UV/IR Mixing\\with a $P(X)$ Higgs Classicalizer}}

\author[a,b]{Eléana Berteaux}
\author[a,1]{and Florian Nortier\note{Corresponding author.}}

\affiliation[a]{Université Lyon 1, CNRS/IN2P3, IP2I Lyon, Villeurbanne, France}
\affiliation[b]{Université de Strasbourg, F-67081 Strasbourg cedex, France}

\emailAdd{eleana.berteaux@etu.univ-lyon1.fr}
\emailAdd{f.nortier@ip2i.in2p3.fr}

\abstract{Classicalization is a non-Wilsonian framework in which an effective field theory self-completes above its strong-coupling scale by generating semiclassical objects that decay into many soft quanta, thereby realizing ultraviolet--infrared mixing. A key requirement of this mechanism is that the force carrier be lighter than the strong-coupling scale. A natural application would be to embed classicalization in the electroweak Higgs sector to address the little hierarchy problem. In this work, we investigate whether Higgs classicalization can occur within the $P(X)$ class of gauged k-essence and k-chameleon models. For classicalization to occur, the theory must support Vainshtein screening around a sharply localized source. Our results show that it fails to do so, even when one attempts to engineer a symbiosis with chameleon screening through a gauge kinetic function, owing to a strong-coupling breakdown inside the screening region. These $P(X)$ theories therefore fail to provide a suitable framework for Higgs classicalization. Nevertheless, we emphasize that our results, obtained for a specific class of models, should not be interpreted as a no-go theorem for Higgs classicalization.}

\begin{document}

\maketitle
\flushbottom

\section{Introduction}
\label{introduction}
The electroweak (EW) hierarchy problem \cite{Susskind:1978ms, tHooft:1979rat, Veltman:1980mj, Kolda:2000wi, Garces:2025rgn}, stemming from the lightness of the Higgs boson, has been central to theoretical model building for nearly 5 decades \cite{Giudice:2008bi,Csaki:2015hcd,Craig:2022eqo}. Following the discovery of a Higgs-like $H$-boson by the CERN LHC experiments \cite{ATLAS:2012yve, CMS:2012qbp, CMS:2022dwd, ATLAS:2022vkf}---yet with no evidence for the new particles predicted by the historical symmetry-based solutions \cite{ParticleDataGroup:2026mpi}---this problem has become even more pressing than it was before the LHC began operating \cite{Cho:2007cb, Dine:2015xga, Giudice:2017pzm, Craig:2020ojv, McCullough:2024evr}.

Motivated by the so-called ``little hierarchy problem'' between the weak scale and the multi-TeV scales of historical approaches, the prospect of a link between ultraviolet (UV) and infrared (IR) scales---one that challenges the traditional Wilsonian picture of effective field theories (EFTs) at the heart of the hierarchy problem \cite{Wilson:1970ag}---has emerged as a possible way to resolve this puzzle\footnote{Alternatives to the approach discussed in this article have been proposed to address the UV sensitivity of scalar masses through some form of UV/IR mixing, including the Swampland program \cite{Cheung:2014vva,Ibanez:2017kvh,Ibanez:2017oqr,Lust:2017wrl,Craig:2019fdy}, misaligned supersymmetry \cite{Dienes:2001se,Abel:2015oxa,Abel:2021tyt,Abel:2024twz}, Lee--Wick theories \cite{Grinstein:2007mp,Espinosa:2007ny,Carone:2008bs,Carone:2008iw,Wise:2009mi,Espinosa:2011js,Brivio:2014pfa,Boos:2021chb}, supergroups \cite{Craig:2024rdg,Craig:2024ypw}, Born-reciprocal tensor networks \cite{Abel:2026gyu}, and infinite-derivative QFTs with \cite{Craig:2019zbn} or without \cite{Krasnikov:1987yj,Moffat:1990jj,Biswas:2014yia,Chattopadhyay:2023nbj} noncommutative geometry.} \cite{Giudice:2017pzm, Craig:2020ojv, McCullough:2024evr}. Such UV/IR mixing is inspired by the anticipated self-completion of gravity in the ultra-Planckian regime via black hole (BH) production \cite{Aharony:1998tt,Giddings:2007qq,Giddings:2009gj,Dvali:2010bf,Dvali:2010ue,Giddings:2011xs,Dvali:2014ila,Addazi:2016ksu}: rather than probing sub-Planckian length scales, as one might naively expect from wave-particle duality in quantum field theory (QFT), one instead produces semiclassical BH-like extended objects that evaporate through Hawking radiation \cite{Hawking:1974rv,Hawking:1975vcx}.


\begin{figure}[t]
\begin{center}
\includegraphics[height=10cm]{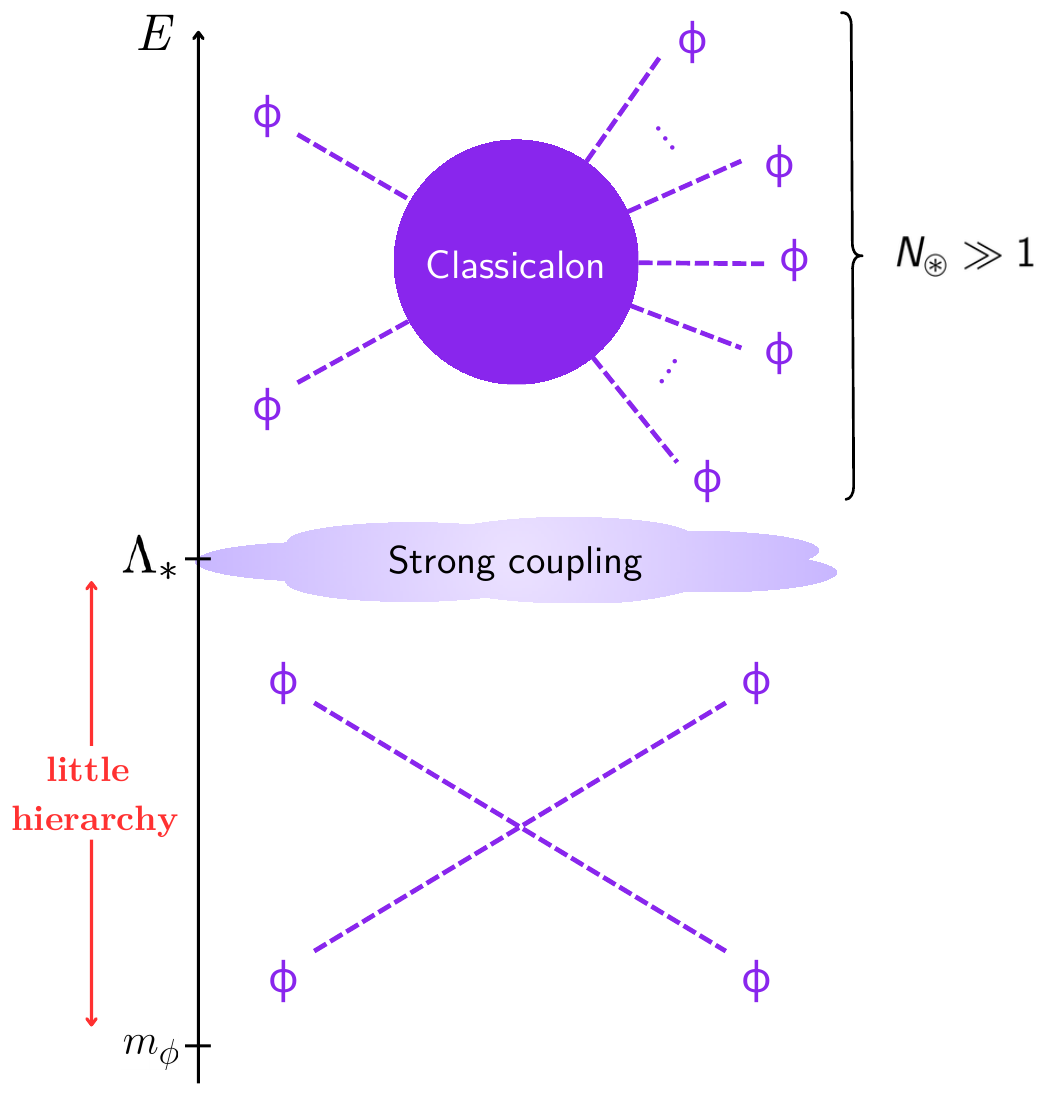}
\end{center}
\caption{
Illustration of self-UV completion via classicalization for a scalar theory with derivative self-interactions set by the strong-coupling scale $\Lambda_\ast$. In a scattering experiment, any attempt to concentrate an energy $E \gg \Lambda_\ast$ in a region of size $1/E$ produces a classicalon whose radius $R_\circledast$ grows with $E$. This extended semiclassical object then decays into $N_\circledast \gg 1$ soft scalars. To evade the strong-coupling breakdown by producing classicalons as coherent states, the theory requires a small hierarchy between the scalar mass $m_\phi$ and $\Lambda_\ast$.
}
\label{Classicalon}
\end{figure}

A concrete field-theoretic realization of UV/IR mixing is the proposal of non-Wilsonian UV completion via classicalization \cite{Dvali:2010jz,Dvali:2010ns,Dvali:2011nj,Dvali:2011th,Dvali:2012zc,Dvali:2016ovn,Nortier:2025gmc}. Rather than integrating in new UV degrees of freedom, the EFT self-completes by reorganizing its IR degrees of freedom into multiparticle states (see Fig.~\ref{Classicalon}), forming extended semiclassical objects (classicalons) that decay into $N_\circledast \gg 1$ soft quanta \cite{Grojean:2011bq,Alberte:2012is,Dvali:2018xoc,Dvali:2022vzz}, in direct analogy with BHs in gravity \cite{Dvali:2011th,Dvali:2016ovn}. Prior to entering this semiclassical regime, the theory crosses a strong-coupling scale $\Lambda_\ast$, at which sharp resonances (fuzzyons) are expected to appear. This scale is associated with the nonperturbative dynamical emergence of a minimal length probed by the classicalizer field.

Classicalizing theories have been characterized as QFTs with nonlocalizable fields \cite{Keltner:2015qqb,Keltner:2015xda,Buoninfante:2023dyd}, which accounts for their violation of positivity bounds \cite{Adams:2006sv,Dvali:2012zc}---bounds derived\footnote{Note that the justification for positivity bounds, which rests on the absence of closed timelike curves on certain field backgrounds~\cite{Adams:2006sv}, fails once quantum corrections are included~\cite{Babichev:2007dw,Burrage:2011cr,Kaplan:2024qtf}.} under the assumption of localizable fields \cite{Tokuda:2019nqb,Buoninfante:2023dyd,Buoninfante:2024ibt}. This situation is not without precedent: gravity---the archetypal classicalizing theory \cite{Dvali:2010bf,Dvali:2010ue,Dvali:2011th}---is well known to exhibit nonlocal features. This is manifest in the ultra-Planckian regime of BH production \cite{Aharony:1998tt,Giddings:2009gj,Giddings:2011xs,Buoninfante:2023dyd}, and is also related to the definition of diffeomorphism-invariant observables in quantum gravity \cite{Torre:1993fq,Giddings:2005id}.




It was established in the literature \cite{Dvali:2010jz,Dvali:2012zc,Brax:2016jjt,Padilla:2017wth} that a fundamental criterion for classicalization is whether the classicalizer field can dress a sharply localized energy-momentum source with a ``cloud'' in which nonlinearities in the field derivatives dominate, thereby preventing field quanta from probing this region \cite{Dvali:2007kt}. This effect is reminiscent of the Vainshtein screening mechanism \cite{Vainshtein:1972sx,Nicolis:2004qq}, widely used in modified gravity to conceal a long-range scalar field from solar-system tests of General Relativity (see Ref.~\cite{Joyce:2014kja} for a review). In Ref.~\cite{Nortier:2025gmc}, it was further shown that a chameleon-like screening mechanism\footnote{As with the Vainshtein mechanism, chameleon screening was proposed to hide scalar dark energy around astrophysical bodies \cite{Khoury:2003aq,Khoury:2003rn,Joyce:2014kja}.}---namely, nonlinearities in the field itself---must supplement Vainshtein-like screening whenever a scalar potential or a nonderivative coupling to the classicalizer is also present, such as a Yukawa coupling to a fermion.

The possibility that classicalization in the Higgs sector could resolve the hierarchy problem was one of the primary motivations for the original proposal of the mechanism \cite{Dvali:2010jz}, which conjectured that a light Higgs boson is favored. This claim was recently reinforced in Ref.~\cite{Nortier:2025gmc}, where it was argued that a little hierarchy between the fluctuation mass of the classicalizer field and $\Lambda_\ast$ is required for the self-completion mechanism to be consistent (see Fig.~\ref{Classicalon}). Given the apparent little hierarchy problem between the mass of the $H$-boson and the next high-energy scale, the prospect of constructing a realistic classicalization model for the EW Higgs boson is particularly compelling.

The present article is a natural follow-up to Ref.~\cite{Nortier:2025gmc}, which provides a more complete bibliography on classicalization and its motivations. Here, we explore the possibility of constructing a Higgs classicalizer within the class of $P(X)$ theories proposed in Ref.~\cite{Dvali:2016ovn}, where $P$ is a polynomial function and $X$ denotes the kinetic term of the Higgs field. $P(X)$ theories form a particular subclass of k-essence theories \cite{Armendariz-Picon:1999hyi,Chiba:1999ka,Armendariz-Picon:2000nqq,Armendariz-Picon:2000ulo,Brax:2014wla,Joyce:2014kja}, which are studied in dark energy phenomenology and are well known to exhibit Vainshtein screening \cite{Brax:2012jr,Joyce:2014kja}. For a Higgs field, the crucial difference relative to the k-essence models of Ref.~\cite{Nortier:2025gmc} is the presence of covariant derivatives involving a gauge field.

This article is organized as follows. In Section~\ref{k-ess_prot}, we review the 2 main building blocks of our discussion: $(i)$ the Vainshtein mechanism in the simplest k-essence model, as a prototype for classicalization \cite{Nortier:2025gmc}; and $(ii)$ the Abelian Higgs model\footnote{See Refs.~\cite{Anderson:1963pc,Higgs:1964ia,Englert:1964et,Guralnik:1964eu,Kibble:1967sv} for historically related articles on the so-called Higgs mechanism.} \cite{Higgs:1964pj, Higgs:1966ev}, emphasizing the features most relevant to the present study. Section~\ref{PX_Abelian_H} presents the original results of this work, in which we show that a $P(X)$ Higgs model cannot realize Vainshtein screening around a sharply localized source, because the nonlinearities cannot prevent the theory from becoming strongly coupled, at which point the field theory ceases to be well defined. We then show that even a k-chameleon model \cite{Wei:2004rw,Nortier:2025gmc,Sirera:2026kzx}, combining Vainshtein and chameleon-like screening \cite{Nortier:2025gmc} through a gauge kinetic function, cannot cure this pathology while preserving a $P(X)$ structure. In Section~\ref{generalization}, we summarize our results and discuss how they generalize, or may be circumvented, in other models of Higgs classicalization to be addressed in future work.

\paragraph{Conventions:}
Unless stated otherwise, our conventions follow those of the QFT textbook by Peskin and Schroeder \cite{Peskin:1995ev}. Let $\phi$ and $A_\mu$ denote scalar and vector fields, respectively. We use the notation $\partial \phi \equiv \pmud \phi$, $(\partial \phi)^{2n} \equiv (\pmuu \phi \pmud \phi)^n$, and $A^{2n} \equiv (A^\mu A_\mu)^n$, with $n \in \mathbb{N}^\ast$.

\section{Preliminaries}
\subsection{\boldmath $P(X)$ K-essence as a Classicalizer Prototype}
\label{k-ess_prot}
In this section, we review the k-essence model analyzed in Ref.~\cite{Nortier:2025gmc}. We consider a real scalar field $\rho(x)$ in $(3+1)$-dimensional Minkowski spacetime with a kinetic self-interaction, coupled to an external (nondynamical) probe source $T(x)$. The Lagrangian density reads
\begin{equation}
\mathcal{L}_\rho \equiv
\frac{1}{2} (\partial \rho)^2
+ \frac{c_2}{4 \Lambda_\ast^4} \left( \partial \rho\right)^4 + \frac{\rho}{\Lambda_\ast} T,
\label{L_rho}
\end{equation}
where $\Lambda_\ast \equiv 1/\ell_\ast$ is the interaction scale and $c_2 = \pm 1$. This toy model has a $\mathbb{Z}_2$ symmetry,
\begin{equation}
\rho(x) \mapsto -\rho(x), \quad
T(x) \mapsto -T(x),
\end{equation}
and a shift symmetry\footnote{Under the shift symmetry, the source term transforms only by adding a constant to the action, since the source $T(x)$ is nondynamical.},
\begin{equation}
\rho(x) \mapsto \rho(x) + \rho_c, \quad
\rho_c \in \mathbb{R}.
\end{equation}
In the same spirit as sources in the path integral formalism \cite{Peskin:1995ev}, $T(x)$ is treated as a spurion transforming as a Lorentz scalar that is odd under the $\mathbb{Z}_2$ symmetry.

In the absence of the probe source ($T(x)=0$), the derivative self-interaction term in Eq.~\eqref{L_rho} enables self-UV completion via classicalization. We emphasize that this nonrenormalizable operator does not arise from integrating out UV degrees of freedom; rather, it represents a fundamental interaction that becomes strong at the scale $\Lambda_\ast$ \cite{Dvali:2010jz,Dvali:2010ns,Dvali:2011nj,Dvali:2011th,Dvali:2012zc,Dvali:2016ovn,Nortier:2025gmc}. For this reason, operators triggering classicalization must be free of Ostrogradsky instabilities \cite{Joyce:2014kja}.

Although this toy model realizes an effectively localized energy-momentum source through $\rho$ boson scattering at center-of-mass energies $\sqrt{s} \gg \Lambda_\ast$ with impact parameter $b \sim 1/\sqrt{s}$, a full analysis of classicalization dynamics in scattering processes remains challenging; numerical studies exist only for idealized cases~\cite{Brouzakis:2011zs,Rizos:2011wj,Rizos:2012qs}. A more tractable approach is to study the response of the classicalizer field $\rho(x)$ to a sharply localized probe source $T(x)$. If a classicalon forms around such a pointlike source, nonlinearities should preclude the localization of hard quanta in realistic scattering processes. To this end, we consider a pointlike source of integrated magnitude $\mathcal{E} \gg \Lambda_\ast$ in its rest frame:
\begin{equation}
T(x) = - \mathcal{E} \delta^{(3)}(\mathbf{r}), \quad
\mathcal{E}>0.
\label{eq:source_0}
\end{equation}
We adopt spherical coordinates $\mathbf{r} \equiv (r,\vartheta,\varphi)$, for which the total solid angle is $\Omega \equiv 4\pi$. The existence of a classicalon solution indicates that the theory is a viable candidate for classicalization. Regarding the source term, once the frame and the sign of $T(x)$ in Eq.~\eqref{eq:source_0} are fixed, both the Lorentz--Poincaré and $\mathbb{Z}_2$ symmetries are broken, while the shift symmetry is preserved.

Integrating the Euler--Lagrange equation for $\rho(r)$ via the divergence theorem yields the cubic equation for $\rho' \equiv d\rho/dr$:
\begin{equation}
\rho' - \frac{c_2}{\Lambda_\ast^4}  (\rho')^3 = \frac{1}{\Omega r^2} \cdot \frac{\mathcal{E}}{\Lambda_\ast}.
\end{equation}
The background solution $\overline{\rho}'(r)$, known as the classicalon, exhibits 3 distinct regimes:
\begin{equation}
\dfrac{\overline{\rho}'(r)}{\Lambda_\ast^2} \simeq
\left\{
\begin{array}{cllll}
\left( \frac{R_V}{r} \right)^2 & \ll 1 & \text{for} & r \gg R_V & \text{(linear regime)}, \\[0.5em]
\mathcal{O}(1) & & \text{for} & r \sim R_V & \text{(transition regime)}, \\[0.5em]
(-c_2)^{1/3} \left( \frac{R_V}{r} \right)^{2/3} & \gg 1 & \text{for} & r \ll R_V & \text{(nonlinear regime)},
\end{array}
\right.
\label{eq:background_real}
\end{equation}
where the transition scale $R_V$ (the Vainshtein radius) is defined as
\begin{equation}
R_V \equiv \ell_\ast \sqrt{\frac{\mathcal{E}}{\Omega \Lambda_\ast}} \gg \ell_\ast.
\end{equation}
Notably, a smooth real solution interpolating between these regimes exists only for $c_2 = -1$, a hallmark of self-UV completion via classicalization~\cite{Dvali:2012zc,Buoninfante:2023dyd,Nortier:2025gmc}. This contrasts with Wilsonian UV completion, which requires $c_2 = +1$~\cite{Pham:1985cr,Adams:2006sv,Dvali:2012zc}.

In the nonlinear regime, the interaction term dominates over the kinetic term, preventing $\rho$-bosons from propagating freely owing to the inertia induced by the kinetic self-interactions. This phenomenon is known as Vainshtein screening~\cite{Vainshtein:1972sx,Nicolis:2004qq}. Zooming in on a point $\mathbf{r}=\mathbf{r}_0$ with resolution $\Delta r \ll r_0$, and neglecting corrections of order $\Delta r/r_0$ and higher, we find that the quantum fluctuations $\delta\rho(x) = \rho(x) - \overline{\rho}(r_0)$ of the classicalon undergo a large field-strength renormalization for $r_0 \ll R_V$; after canonical normalization, the interaction scale inside the classicalon, $\Lambda_B(r_0)$, is blueshifted:
\begin{equation}
\delta\rho_Z(x) \equiv \sqrt{Z(r_0)}  \delta\rho(x),
\quad
\Lambda_B(r_0) \equiv \sqrt{Z(r_0)}  \Lambda_\ast,
\label{eq:blueshift}
\end{equation}
where the renormalization factor is
\begin{equation}
Z \left( r_0 \ll R_V \right) \simeq \left[\frac{\overline{\rho}'(r_0)}{\Lambda_\ast^2}\right]^2  \gg 1.
\label{eq:Z_factor}
\end{equation}
The leading cubic and quartic interaction terms are
\begin{equation}
\dfrac{(\partial_r \delta\rho_Z) (\partial \delta\rho_Z)^2}{\Lambda_B^2}
- \dfrac{(\partial \delta\rho_Z)^4}{4\Lambda_B^4}
+ \mathcal{O} \left( \frac{\Delta r}{r_0} \right).
\end{equation}
Consequently, the background solution with $\overline{\rho}'(r_0 \ll R_V) \gg \Lambda_\ast^2$ remains valid, even though it appears at first glance to lie outside the EFT regime of validity. Since $\Lambda_B(r_0)$ is the local interaction scale, it also governs the quantum corrections around the background solution at $\mathbf{r} = \mathbf{r}_0$ \cite{Nicolis:2004qq}. The closer one gets to the source, the more suppressed these quantum corrections become \cite{Nicolis:2004qq,Vikman:2012bx,Brouzakis:2014bwa,deRham:2014wfa,Asimakis:2014bza,Padilla:2017wth}.

As a final remark, the Lagrangian density of the fluctuations is not Lorentz--Poincaré invariant. This is expected, since we expand around a background that breaks Lorentz--Poincaré symmetry: the background solution \eqref{eq:background_real} is spherically symmetric about its source and invariant under time translations only. This feature will be shared by all $P(X)$ models studied in this article.

\subsection{Abelian Higgs Model}
\label{Ab_H_Model}
In this section, we review the basic features of the Abelian Higgs model \cite{Higgs:1964pj, Higgs:1966ev}---a $U(1)$ gauge theory---which serves as a toy model for giving mass to a photon-like particle. On $(3+1)$-dimensional Minkowski spacetime, we consider a complex scalar field $\phi(x)$ of charge $-1$ and potential $V(\phi)$, minimally coupled to a photon field $A_\mu(x)$ with Faraday tensor\footnote{
For a rank-2 tensor $\mathcal{T}_{\mu\nu}$, the antisymmetrization operation is denoted
\begin{equation}
\mathcal{T}_{[\mu\nu]} \equiv \frac{\mathcal{T}_{\mu\nu} - \mathcal{T}_{\nu\mu}}{2}.
\end{equation}
} $F_{\mu\nu} = 2  \partial_{[\mu} A_{\nu]}$. The Lagrangian density reads
\begin{equation}\label{eq:Lagr higgs}
\lagr_\mathrm{Higgs} \equiv -\frac{1}{4}  F_{\mu\nu}F^{\mu\nu} + (D_\mu \phi)^* (D^\mu \phi) - V(\phi),
\end{equation}
which is invariant under the $U(1)$ gauge transformations:
\begin{equation}\label{eq:U(1) transfo}
\phi(x) \mapsto e^{ig\alpha(x)} \phi(x), \quad
A_\mu(x) \mapsto A_\mu(x) - \pmud \alpha(x), \quad
\alpha(x) \in \mathbb{R},
\end{equation}
with the covariant derivative $D_\mu = \pmud + ig  A_\mu$ and the gauge coupling $g$.

For a renormalizable Abelian Higgs model, the potential takes the form
\begin{equation}\label{eq:Higgs pot}
V(\phi) = \mu^2  \phi^* \phi + \lambda (\phi^* \phi)^2,
\end{equation}
with $\mu^2 < 0$ and $\lambda > 0$. Consequently, the Higgs field $\phi(x)$ acquires a vacuum expectation value (VEV), and the $U(1)$ symmetry is spontaneously broken. The potential is minimized at
\begin{equation}\label{eq:Higgs pot min}
v \equiv \left\langle \phi(x) \right\rangle = \sqrt{-\frac{\mu^2}{\lambda}},
\end{equation}
and the infinitely many degenerate minima are connected by the continuous $U(1)$ transformation of $\phi(x)$ in Eq.~\eqref{eq:U(1) transfo}.

Below, we discuss the Lagrangian density for fluctuations about the VEV in both the polar and linear parameterizations of the Higgs field $\phi(x)$.

\paragraph{Polar Parameterization:}
The polar parameterization of the Higgs field is given by
\begin{equation}\label{eq:polar param}
    \phi(x) \equiv \frac{\rho(x)}{\sqrt{2}}  e^{i\theta(x)},
\end{equation}
where $\rho(x)>0$ and $\theta(x)$ are the real ``radial'' and ``phase'' fields, respectively. Applying the $U(1)$ transformation \eqref{eq:U(1) transfo}, we can choose the unitary gauge
\begin{equation}
\phi(x) \mapsto e^{ig\alpha(x)}  \phi(x), \quad
\alpha(x) = -\frac{\theta(x)}{g},
\label{gauge_transfo}
\end{equation}
in which the Higgs field is real: $\phi(x) = \rho(x) / \sqrt{2}$. Eq.~\eqref{eq:Lagr higgs} then reduces to
\begin{equation}\label{eq:Lagr higgs unitary gauge}
\lagr_\mathrm{Higgs} = -\frac{1}{4}  F^{\mu\nu}F_{\mu\nu} + \half  (\partial \rho)^2 + \dfrac{g^2}{2}  \rho^2 A^2  - \dfrac{\mu^2}{2}  \rho^2 - \frac{\lambda}{4}  \rho^4.
\end{equation}
In terms of the fluctuation field about the VEV,
$\rho(x) = v + \delta\rho(x)$, the Lagrangian density reads
\begin{align}
    \lagr_\mathrm{Higgs} &= \half  (\partial \delta\rho)^2 - \dfrac{M_H^2}{2} \delta\rho^2 -\frac{1}{4}  F^{\mu\nu}F_{\mu\nu} + \dfrac{M_A^2}{2}  A^2 \nonumber\\
    &\quad + g^2 v  (\delta\rho) A^2 + \dfrac{g^2}{2}  (\delta\rho)^2 A^2 - \lambda v  (\delta\rho)^3 - \dfrac{\lambda}{4}  (\delta\rho)^4,
\end{align}
which describes a massive scalar field $\delta\rho(x)$ (the Higgs boson) and a massive photon $A_\mu(x)$, with masses
\begin{equation}
     M_H \equiv \sqrt{2 \lambda} v \quad \text{and} \quad M_A \equiv gv,
\end{equation}
respectively. It also includes interactions between the Higgs boson and the photon, as well as Higgs self-interactions. In unitary gauge, the unphysical phase field $\theta(x)$---the would-be Goldstone boson of spontaneous symmetry breaking---is eaten by the photon, which thereby acquires mass and supplies the longitudinal polarization.

\paragraph{Linear Parameterization:}
We now consider the linear parameterization of the Higgs field:
\begin{equation}\label{eq:lin param}
    \phi(x) \equiv \frac{1}{\sqrt{2}}[\phi_1(x) + i\phi_2(x)],
\end{equation}
where $\phi_1(x)$ and $\phi_2(x)$ are real fields. As before, expanding the Higgs field about its VEV as $\phi_1(x)=v+\delta\phi_1(x)$ and $\phi_2(x)=\delta\phi_2(x)$ gives
\begin{align}
\lagr_\mathrm{Higgs} &\supseteq \frac{1}{2}  (\partial \delta\phi_1)^2 - \dfrac{M_H^2}{2}  (\delta\phi_1)^2 + \frac{1}{2}  (\partial \phi_2)^2 \nonumber\\
& - \frac{1}{4}  F^{\mu\nu}F_{\mu\nu} + \dfrac{M_A^2}{2}  A^2 + gv  A_\mu \partial^\mu \phi_2,
\end{align}
where the interaction terms are not shown. The unitary gauge is obtained by the gauge transformation
\begin{equation}
A_{\mu} (x) \mapsto A_{\mu}(x) - \partial_\mu \alpha(x), \quad
\alpha(x) = \frac{\phi_2(x)}{gv},
\label{gauge_transfo_inf}
\end{equation}
which removes the would-be Goldstone boson $\phi_2(x)$ from the quadratic Lagrangian density. Note that $\phi_2(x)$ can be identified with the parameter of the gauge transformation in Eq.~\eqref{gauge_transfo_inf} only when it is infinitesimal, i.e., $\phi_2(x) \ll gv$: comparing with the polar parameterization for $\theta(x)\ll 1$ in Eqs.~\eqref{eq:polar param}-\eqref{gauge_transfo},
\begin{equation}
\phi(x) \simeq \dfrac{1}{\sqrt{2}} \left[ v+\delta\rho(x) + iv  \theta(x) \right] + \mathcal{O}( \theta^2 ),
\end{equation}
yields the identifications $\delta\phi_1(x)=\delta\rho(x)$ and $\phi_2(x) = v  \theta(x)$. This truncation explains why the gauge transformation \eqref{gauge_transfo_inf} removes $\phi_2(x)$ from the quadratic Lagrangian density but not from the interaction terms. A more systematic treatment of unitary gauge in the linear parameterization is obtained by introducing gauge-fixing terms motivated by the $R_\xi$ gauge ($\xi \to \infty$) in the path integral quantization of the theory \cite{tHooft:1971qjg,Peskin:1995ev}.

\section{\boldmath $P(X)$ Higgs Model}
\label{PX_Abelian_H}

\subsection{\boldmath $P(X)$ Generalization of the Abelian Higgs Model}
We now promote the usual Abelian Higgs model (Section~\ref{Ab_H_Model}) to a gauge-covariant $P(X)$ k-essence model (Section~\ref{k-ess_prot}). To this end, we define the kinetic function
\begin{equation}
P_n(X) \equiv X + c_n X^n, \quad n \in \{ 2, 3, 4, \cdots \},
\end{equation}
with no summation over $n$, and $c_n \equiv \pm 1$. The kinetic variable $X$ is defined as
\begin{equation}
        X \equiv \frac{(D^\mu\phi)^* (D_\mu \phi)}{\Lambda_\ast^4}
        = \frac{1}{\Lambda_\ast^4} \left[ \pmuu\phi^* \pmud \phi + g j^\mu A_\mu + g^2 \phi^* \phi A^2 \right],
    \label{eq:X der cov}
\end{equation}
where the scalar current is
\begin{equation}\label{eq:scalar_current}
j^\mu \equiv i\left( \phi \partial^\mu \phi^* - \phi^*\partial^\mu \phi \right).
\end{equation}
The Abelian $P(X)$ Higgs model is described by the Lagrangian density
\begin{equation}
\mathcal{L}_X \equiv -\frac{1}{4}  F^{\mu\nu} F_{\mu\nu} + \Lambda_\ast^4  P_n(X) - V(\phi) + \frac{\phi^\ast}{\Lambda_\ast}  T + \frac{\phi}{\Lambda_\ast}  T^\ast,
\label{eq:PX_Lag_0}
\end{equation}
where the potential is left unspecified but must be chosen so as to realize spontaneous symmetry breaking. We have also included source terms, in which the (nondynamical) probe source $T(x)$ is a spurion that must transform in the same way as $\phi(x)$, so that the $U(1)$ and Lorentz--Poincaré symmetries are preserved. Appendix~\ref{Lag_exp} presents the expansion of this Lagrangian density in powers of $g$ for $n=2$ and $c_2=-1$.

We aim to perform a study in the same spirit as Section~\ref{k-ess_prot} and Ref.~\cite{Nortier:2025gmc}. As we will see, the presence of the sources alone induces a Higgs-like mechanism, and the potential will not be crucial for our negative conclusions regarding Vainshtein screening; we therefore drop it for the moment, setting $V(\phi)=0$. In addition, we focus on the simplest model proposed in Ref.~\cite{Dvali:2016ovn}, with $n=2$, and we discuss how our conclusions generalize to $n>2$ in Section~\ref{generalization}.

Before proceeding, we point out that this $P(X)$ Higgs model \eqref{eq:PX_Lag_0}, proposed in Ref.~\cite{Dvali:2016ovn} (with $n=2$), is not the first proposal for a Higgs classicalizer. Indeed, a different derivative self-interaction operator was suggested in Ref.~\cite{Dvali:2010jz}, whose collider phenomenology was explored in Ref.~\cite{Grojean:2011bq}. In Appendix~\ref{fate_1st_model}, we explain why that possibility is no longer considered for realizing self-UV completion by classicalization.

\subsection{Screening in the Decoupling Limit}
\label{decoupling_limit}
\subsubsection{\boldmath Complex $P(X)$ K-essence}
In the decoupling limit $g \to 0$, the photon decouples from the scalar field and thus from the classicalization dynamics, so we may set $A_\mu \to 0$. The Lagrangian density \eqref{eq:PX_Lag_0} reads
\begin{equation}
    \lagr_X = \pmuu\phi^* \pmud\phi + \frac{ c_2 }{\Lambda_\ast^4} \left(\pmuu\phi^* \pmud\phi \right)^2 + \frac{\phi^\ast}{\Lambda_\ast} T + \frac{\phi}{\Lambda_\ast} T^\ast,
    \label{eq:LX NR}
\end{equation}
with $n=2$ and $V(\phi) = 0$.
Unlike the real scalar of Section~\ref{k-ess_prot}, which has $\mathbb{Z}_2$ and real shift symmetries, the complex scalar field has a global $U(1)$ symmetry,
\begin{equation}
\phi(x) \to e^{i\beta} \phi(x), \quad
T(x) \to e^{i\beta} T(x), \quad
\beta \in \mathbb{R},
\end{equation}
and a complex shift symmetry,
\begin{equation}
\phi(x) \mapsto \phi(x) + \phi_c, \quad
\phi_c \in \mathbb{C}.
\label{eq:complex_shift}
\end{equation}

To study this complex field, we use the linear parameterization \eqref{eq:lin param}, in terms of which the Lagrangian density \eqref{eq:LX NR} becomes
    \begin{equation}
         \lagr_X = \sum_{i=1}^2 \left[ \half (\partial \phi_i)^2 + \frac{c_2}{4 \Lambda_\ast^4} (\partial \phi_i)^4 \right] + \frac{c_2}{2 \Lambda_\ast^4} (\partial \phi_1)^2(\partial \phi_2)^2,
         \label{eq:Lag_12}
    \end{equation}
with $T(x)=0$.
A new feature of the complex scalar field is that the nonlinear derivative self-interactions couple its real and imaginary components through the last term of the Lagrangian density.

We now determine how the complex field $\phi(x)$ responds to the external source
\begin{equation}
T(x) = \frac{1}{\sqrt{2}}\left[T_1(x) + iT_2(x)\right],
\end{equation}
which we take to be pointlike and express in its rest frame as
\begin{equation}
    T = -\frac{1}{\sqrt{2}}(\mc{E}_1 + i\mc{E}_2) \delta^{(3)}(\mathbf{r}), \quad \text{with} \quad \int d^3 \mathbf{r} \ \delta^{(3)}(\mathbf{r})=1.
\end{equation}
This is a sharply localized complex source whose real and imaginary parts, $T_1(x)$ and $T_2(x)$, have positive magnitudes $\mathcal{E}_1$ and $\mathcal{E}_2$, respectively. Since $\phi(x)$ is complex, we can write it in terms of 2 real fields as in \eqref{eq:lin param}; the source terms in Eq.~\eqref{eq:LX NR} then read
\begin{equation}\label{eq:LJ}
\frac{\phi^\ast}{\Lambda_\ast} T + \frac{\phi}{\Lambda_\ast} T^\ast
= -\frac{1}{\Lambda_\ast} \left( \mc{E}_1 \phi_1 + \mc{E}_2 \phi_2 \right) \delta^{(3)} (\mathbf{r}).
\end{equation}
Once the frame and the phase of the complex source are fixed, the source terms break the Lorentz--Poincaré and global $U(1)$ symmetries while preserving the shift symmetry of the action.

\subsubsection{Scalar Background}
The Euler--Lagrange equations for $\phi_i$, $i=1,2$, in the presence of their sources read
\begin{equation}\label{eq:EL phi1}
\square \phi_i + \frac{c_2}{\Lambda_\ast^4}
\partial_\mu \left[ \partial^\mu \phi_i \sum_{j=1}^2 \left( \partial \phi_j \right)^2 \right] = - \frac{\mathcal{E}_i}{\Lambda_\ast} \delta^{(3)}(\mathbf{r}).
\end{equation}
Exploiting the spherical symmetry of the static source, the divergence theorem yields the following cubic equations for the fields:
\begin{align}\label{eq:Gauss}
\left[ 1 - \frac{c_2}{\Lambda_\ast^4} \sum_{j=1}^2 (\phi_j^\prime)^2 \right] \phi_i^\prime = \frac{1}{\Omega r^2} \cdot \frac{\mathcal{E}_i}{\Lambda_\ast}.
\end{align}
If $\mathcal{E}_i=0$ for $i=1$ or $2$ but not for the other index, then $\phi_i^\prime = 0$, and the problem reduces to the Vainshtein screening of a single scalar field discussed in Section~\ref{k-ess_prot}. In what follows, we therefore focus on the more interesting case in which both sources are nonzero. The 2 equations can be decoupled as
\begin{equation}\label{final eq}
    \phi_i' - \frac{c_2}{\Lambda_\ast^4} \left( \frac{\mc{E}_{12}}{\mc{E}_i} \right)^2 (\phi_i')^3 = \frac{1}{\Omega r^2} \cdot \frac{\mathcal{E}_i}{\Lambda_\ast},
\end{equation}
with $\mc{E}_{12} \equiv \sqrt{\mc{E}_1^2 + \mc{E}_2^2} \gg \Lambda_\ast$. This cubic equation in $\phi'_i$ admits a solution that allows us to identify 3 distinct regimes: the linear regime, where $\phi'_i \gg \ell_\ast^4 (\mc{E}_{12}/\mc{E}_i)^2 \phi'_i{}^3$; the nonlinear regime, where $\phi'_i \ll \ell_\ast^4 (\mc{E}_{12}/\mc{E}_i)^2 \phi'_i{}^3$; and the transition regime at the Vainshtein radius $R_V$, where nonlinearities begin to matter and both terms on the left-hand side of Eq.~\eqref{final eq} are of the same order of magnitude. Hence,
\begin{equation}
    R_V \equiv \ell_\ast\sqrt{\frac{\mc{E}_{12}}{\Omega\Lambda_\ast}}.
\end{equation}
Remarkably, there is only 1 Vainshtein radius, not 2 (1 for each real field). A smooth background solution $\overline{\phi_i}'(r)$ exists only for $c_2=-1$; in the 3 regimes it is given by
\begin{equation}\label{regimes}
    \frac{\overline{\phi}_i'}{\Lambda_\ast^2} \simeq
    \begin{cases}
        \left( \frac{\mc{E}_i}{\mc{E}_{12}} \right) \left( \frac{R_V}{r} \right)^2 \ll 1 & \text{for } r \gg R_V \text{ (linear regime),} \\[10pt]
        \mc{O}(1) & \text{for } r \sim R_V \text{ (transition regime),} \\[10pt]
        \left( \frac{\mc{E}_i}{\mc{E}_{12}} \right) \left( \frac{R_V}{r} \right)^{2/3} \gg 1 & \text{for } r \ll R_V \text{ (nonlinear regime).}
    \end{cases}
\end{equation}
In the limit $\mathcal{E}_i \to 0$ for 1 of the 2 sources, one recovers the situation of Section~\ref{k-ess_prot}, with a nontrivial background only for the field $\phi_{j \neq i}$, sourced by $\mc{E}_j=\mc{E}_{12} \gg \Lambda_\ast$. Given our previous remark on the vanishing-source case, the limit $\mathcal{E}_i \to 0$ is smooth.

\subsubsection{Scalar Fluctuations}
Because the screening region is dominated by nonrenormalizable operators, it is important to check that the fluctuations $\delta \phi_i(x) \equiv \phi_i(x) - \overline{\phi}_i(r)$ about the backgrounds do not spoil the semiclassical analysis. We distinguish the case in which 1 of the 2 sources vanishes ($\mathcal{E}_1 \gg \Lambda_\ast$ and $\mathcal{E}_2=0$; the case $\mathcal{E}_1=0$ follows by exchanging the indices $1 \leftrightarrow 2$) from the case in which both are active ($\mathcal{E}_1, \mathcal{E}_2 \gg \Lambda_\ast$). In the rest of this article, whenever we refer to Vainshtein screening in the decoupling limit, only the single-source case ($\mathcal{E}_2 = 0$) will be considered.

\paragraph{\boldmath 1 Source ($\mathcal{E}_1 \gg \Lambda_\ast$, $\mathcal{E}_2=0$):}
We consider the fluctuations $\delta \phi_1(x) = \phi_1(x)-\overline{\phi}_1(r)$ and $\delta \phi_2(x) = \phi_2(x)$ in the Lagrangian density \eqref{eq:Lag_12}, whose kinetic and interaction terms read
\begin{equation}
\mathcal{L}_\mathrm{kin1} \equiv
- \left( \frac{\overline{\phi}_1^\prime}{\Lambda_\ast^2} \right)^2 (\partial_r \delta \phi_1)^2
+ \left[ 1 + \left(\frac{\overline{\phi}_1^\prime}{\Lambda_\ast^2} \right)^2 \right] \sum_{i=1}^2 \frac{(\partial \delta \phi_i)^2}{2},
\label{eq:Lkin1}
\end{equation}
and
\begin{equation}
\mathcal{L}_\mathrm{int1} \equiv
- \frac{(\partial \delta \phi_{1})^2 (\partial \delta \phi_{2})^2}{2\Lambda_\ast^4}
+ \sum_{i=1}^2 \left[ \frac{\overline{\phi}_1^\prime}{\Lambda_\ast^4} (\partial_r \delta \phi_{1})(\partial \delta \phi_{i})^2 - \frac{(\partial \delta \phi_{i})^4}{4\Lambda_\ast^4} \right],
\label{eq:Lint1}
\end{equation}
respectively. Focusing on the nonlinear regime, where $\overline{\phi}_1^\prime(r_0 \ll R_V) \gg \Lambda_\ast^2$, we zoom in on a point $\mathbf{r}=\mathbf{r}_0$ at resolution $\Delta r \ll r_0$ and neglect corrections of order $\Delta r/r_0$ and higher. We perform the field-strength renormalization
\begin{equation}
\delta \phi_{iZ}(x) = \sqrt{Z_i(r_0)} \delta \phi_{i}(x), \quad
Z_i(r_0) \simeq \left[ \frac{\overline{\phi}_1^\prime(r_0)}{\Lambda_\ast^2} \right]^2,
\end{equation}
to canonically normalize the kinetic terms as
\begin{equation}
\mathcal{L}_\mathrm{kin1} \simeq
- (\partial_r \delta \phi_{1Z})^2
+ \sum_{i=1}^2 \frac{(\partial \delta \phi_{iZ})^2}{2}
+ \mathcal{O}\left( \frac{\Delta r}{r_0} \right),
\end{equation}
such that the interaction terms become
\begin{equation}
\mathcal{L}_\mathrm{int1} \simeq
- \frac{(\partial \delta \phi_{1Z})^2 (\partial \delta \phi_{2Z})^2}{2\Lambda_{B,i}^4} +
\sum_{i=1}^2 \left[ \frac{(\partial_r \delta \phi_{1Z})(\partial \delta \phi_{iZ})^2}{\Lambda_{B,i}^2} - \frac{(\partial \delta \phi_{iZ})^4}{4\Lambda_{B,i}^4} \right] + \mathcal{O} \left( \frac{\Delta r}{r_0} \right),
\end{equation}
with the same effective interaction scale
\begin{equation}
\Lambda_{B,i}(r_0) \equiv \sqrt{Z_i(r_0)} \Lambda_\ast,
\end{equation}
which is blueshifted for all fluctuations, indicating that the EFT remains robust in the nonlinear regime. The real scalar case of Section~\ref{k-ess_prot} and Ref.~\cite{Nortier:2025gmc} is recovered by setting $\delta \phi_{2Z} \to 0$ and identifying $\phi_1(x)$ with $\rho(x)$.

\paragraph{\boldmath 2 Sources ($\mathcal{E}_1, \mathcal{E}_2 \gg \Lambda_\ast$):}
In this case, a gradient mixing between $\delta \phi_1$ and $\delta \phi_2$ arises through the term $(\partial_r \delta \phi_1) (\partial_r \delta \phi_2)$ induced when both backgrounds $\overline{\phi}_1^\prime(r)$ and $\overline{\phi}_2^\prime(r)$ are present. Since it will not be needed for the $P(X)$ Higgs model, the main focus of this article, we do not discuss it further here; it is nonetheless clear that the blueshift effect still operates.

\subsubsection{Chameleon Potential}
Given the success of Vainshtein screening in the decoupling limit, one may now turn on the potential and ask whether any constraints on its form are needed to preserve Vainshtein screening. The discussion parallels that for a real scalar field in Ref.~\cite{Nortier:2025gmc}, where we found that a potential flattening at large values of the classicalizer satisfies this requirement. In the case of our complex field, one can take
\begin{equation}
V(\phi) = \Lambda_\phi^4 \tanh \left[ \dfrac{ \lambda \left( |\phi|^2 - \frac{v^2}{2} \right)^2}{\Lambda_\phi^4} \right] \simeq
\begin{cases}
\mu^2 |\phi|^2 + \lambda |\phi|^4 + \frac{\mu^4}{4\lambda} & \text{for } |\phi| \ll \Lambda_\phi, \\[10pt]
        \Lambda_\phi^4 & \text{for } |\phi| \gg \Lambda_\phi,
    \end{cases}
    \label{eq:}
\end{equation}
with a new scale $\Lambda_\phi \sim \Lambda_\ast$, where $v$ is defined as in Eq.~\eqref{eq:Higgs pot min}. Thus, the potential \eqref{eq:Higgs pot} is recovered for $|\phi| \ll \Lambda_\ast$, while the potential flattens for $|\phi| \gg \Lambda_\ast$. Because this behavior relies on nonlinearities in $\phi$, triggered by the sources that realize Vainshtein screening through the derivative self-interaction, this constitutes chameleon-like screening: the field adapts to its environment \cite{Khoury:2003aq,Joyce:2014kja}. This also exemplifies a symbiosis between chameleon and Vainshtein screening \cite{Nortier:2025gmc,Sirera:2026kzx}.

Regarding the fluctuations $\delta \phi_i(x)$ about the backgrounds $\overline{\phi}_i(r)$ ($i=1,2$) in Eq.~\eqref{regimes}, one can check that they do not spoil the screening mechanism. Indeed, we have the asymptotic behavior:
\begin{equation}
\tanh \left( \lambda \left| \frac{\phi}{\Lambda_\phi} \right|^4 \right)
\underset{|\phi| \gg \Lambda_\phi}{\simeq}
1-2 e^{-2 \lambda \left| \frac{\phi}{\Lambda_\phi} \right|^4},
\end{equation}
valid in the screening region. Hence, expanding this function in the perturbations $\delta \phi_i(x)$ generates polynomial terms involving products of these fluctuations with the backgrounds $\overline{\phi}_i$, all carrying the common prefactor:
\begin{equation}
-2 e^{-2 \lambda \left( \frac{\overline{\phi}_1^2 + \overline{\phi}_2^2}{\Lambda_\phi^2} \right)^2},
\end{equation}
which dominates any polynomial dependence, even when 1 of the 2 backgrounds vanishes (no source), so that the vertices are strongly suppressed in the screening region. The mechanism is the same as that discussed in Ref.~\cite{Nortier:2025gmc}.

\subsection{\boldmath Fate of the $P(X)$ Higgs Classicalizer}
\subsubsection{Strong Coupling Breakdown of Vainshtein Screening}
\label{strong_break_V}
We now turn to the case that motivates our analysis: a perturbative, nonzero gauge coupling $g$ that explicitly breaks the complex shift symmetry \eqref{eq:complex_shift}. Unlike in the decoupling case, 1 of the 2 scalar degrees of freedom is unphysical once the Higgs mechanism occurs: it is eaten by the photon, which thereby acquires a longitudinal polarization and becomes massive. Consequently, the limit $g \to 0$ is not smooth in unitary gauge, but this is merely a gauge artifact.

We work in the polar parameterization of both the Higgs field \eqref{eq:polar param} and the probe source:
\begin{equation}
T(x) = |T(x)| e^{i \cdot \mathrm{arg}[T(x)]}.
\end{equation}
The latter is a spurion under the $U(1)$ gauge symmetry. Therefore, only the phase difference $\Delta \theta(x) \equiv \theta(x) - \mathrm{arg}[T(x)]$ is physical. We are interested in the classicalization of the Higgs boson, so we seek a source coupled only to the radial field, achieved by choosing $\Delta \theta(x) = 0$; the source term then reads
\begin{equation}\label{eq:LJ2}
\frac{\phi^\ast}{\Lambda_\ast} T + \frac{\phi}{\Lambda_\ast} T^\ast
= -\frac{\rho}{\Lambda_\ast} \mathcal{E} \delta^{(3)} (\mathbf{r}), \quad \mathcal{E}>0.
\end{equation}
Going to unitary gauge in the Lagrangian density \eqref{eq:PX_Lag_0} with $n=2$, the Higgs field becomes real, $\phi(x) = \rho(x)/\sqrt{2}$; hence $j^\mu = 0$ from Eq.~\eqref{eq:scalar_current}, and the terms with odd powers of $g$ in Appendix~\ref{Lag_exp} vanish.

Assuming perturbativity in $g$, we determine the background $\overline{\rho}(r)$ at order $g^0$, ignoring all couplings to $A_\mu$; the discussion thus reduces to that of Section~\ref{k-ess_prot}. We then seek the Lagrangian density of the fluctuations $\delta \rho(x) = \rho(x)-\overline{\rho}(r_0)$. Again at order $g^0$, the analysis follows Section~\ref{k-ess_prot}, and the blueshifted interaction scale can be defined after the field-strength redefinition of the fluctuations $\delta \rho_Z$, as in Eq.~\eqref{eq:blueshift}.

The crucial difference from Section~\ref{k-ess_prot} is, of course, the gauge coupling of the scalar background and its fluctuations to the photon. In Appendix~\ref{fluct_int_terms}, we display the Lagrangian density for the nontrivial powers of $g$. The main obstructions to Vainshtein screening can be illustrated by picking 2 terms inside the classicalon ($\ell_\ast \ll r_0 \ll R_V$):
\begin{itemize}[label=$\spadesuit$]
\item Through its interaction with the background, the photon acquires effective couplings, such as the 4-photon coupling term,
\begin{equation}
-g^4 \underbrace{\left( \frac{3r_0}{\ell_\ast} \right)^4 \frac{Z^2}{4}}_{\gg 1} A^4,
\end{equation}
from Eq.~\eqref{eq:Lint 4 non lin regime}, which is driven far beyond perturbative control inside the nonlinear region ($Z \gg 1$) even for a moderately weak coupling $g \lesssim 1$. Moreover, the other couplings between $A_\mu$ and $\delta \rho_Z$ in Eq.~\eqref{eq:Lint 4 non lin regime} show that no blueshifted effective scale can be defined for them.
\item This becomes clear when one realizes that, for the same reasons, the photon acquires an effective mass $M_A^{\mathrm{eff}}$, such that
\begin{equation}
    \dfrac{M_A^{\mathrm{eff}}}{g} \simeq \underbrace{\left( \frac{3r_0}{\ell_\ast} \right)\sqrt{Z}}_{\gg 1} \Lambda_B \gg \Lambda_B,
\end{equation}
from Eq.~\eqref{eq:m eff non lin regime}. This is a Higgs mechanism triggered by the $U(1)$ breaking induced by the source $T(\mathbf{r})$. The problem is that, for a moderately weak coupling $g \lesssim 1$, $M_A^{\mathrm{eff}} \gg \Lambda_B$: the massive photon cannot be treated within the EFT on top of the background, even though Vainshtein screening is supposed to preserve the validity of the original EFT when nonlinearities dominate.
\end{itemize}
These issues can be mitigated for a very suppressed coupling\footnote{Indeed, as discussed in Section~\ref{decoupling_limit}, there is no obstruction to Vainshtein screening in the decoupling limit $g \to 0$.} $g \lll 1$, but this is unsuitable if the goal is to generalize such a model to the EW sector later. We therefore conclude that Vainshtein screening cannot restore the validity of the EFT inside the region where nonlinearities dominate, so the model cannot realize Higgs classicalization\footnote{Our results show that the statement in Section~4 of Ref.~\cite{Dvali:2016ovn}---``\textit{The gauge structure, which plays no role in this discussion, has been ignored}''---made in order to focus solely on the scalar sector of the $P(X)$ Higgs model, was too naive, since the gauge structure is precisely what destroys the consistency of the EFT inside the classicalon. This statement is correct in the context of backgrounds in traditional EFTs that are weakly coupled, as at the exterior of the classicalon ($\overline{\rho}' \ll \Lambda_\ast^2$), but should be treated with care in the context of Vainshtein screening.}, irrespective of the presence of a potential $V(\phi)$.

As a cross-check, instead of the polar parameterization \eqref{eq:polar param} of the Higgs field, we have repeated the analysis with the linear one \eqref{eq:lin param}; we give only the outline here. In that case, we sourced only the $\phi_1$ field (Eq.~\eqref{eq:LJ} with $\mathcal{E}_2 = 0$), whose fluctuations provide the Higgs mode. We studied these fluctuations about the background \eqref{regimes} in the nonlinear region: $\delta \phi_1(x) \equiv \phi_1(x) - \overline{\phi_1}(r)$ and $\delta \phi_2(x) \equiv \phi_2(x)$, zooming in on the point $\mathbf{r} \equiv \mathbf{r_0}$ and thereby dropping corrections of order $\Delta r / r_0$. To go to unitary gauge, one can perform the gauge transformation:
\begin{equation}
A_\mu(x) \mapsto A_\mu(x) - \partial_\mu \alpha(x), \quad
\alpha(x) = \dfrac{\phi_2(x)}{g \overline{\phi_1}(r_0)},
\end{equation}
which is equivalent to Eq.~\eqref{gauge_transfo_inf} with the replacement $v \mapsto \overline{\phi_1}(r_0)$. This simply reflects that we expanded locally around the tangent plane to the background, dropping derivative corrections. We obtain a quadratic action independent of the Goldstone mode $\delta\phi_2(x)$, which is eaten by the massive photon. Given the subtleties of external sources when both physical and spurious degrees of freedom are present in a gauge theory, this cross-check confirms that our results in unitary gauge are not an artifact of the Higgs parameterization.

\subsubsection{\boldmath No Rescue with Chameleon--Vainshtein Symbiosis}
\label{no_rescue}
\begin{figure}[t]
    \centering
    \includegraphics[height=7.8cm]{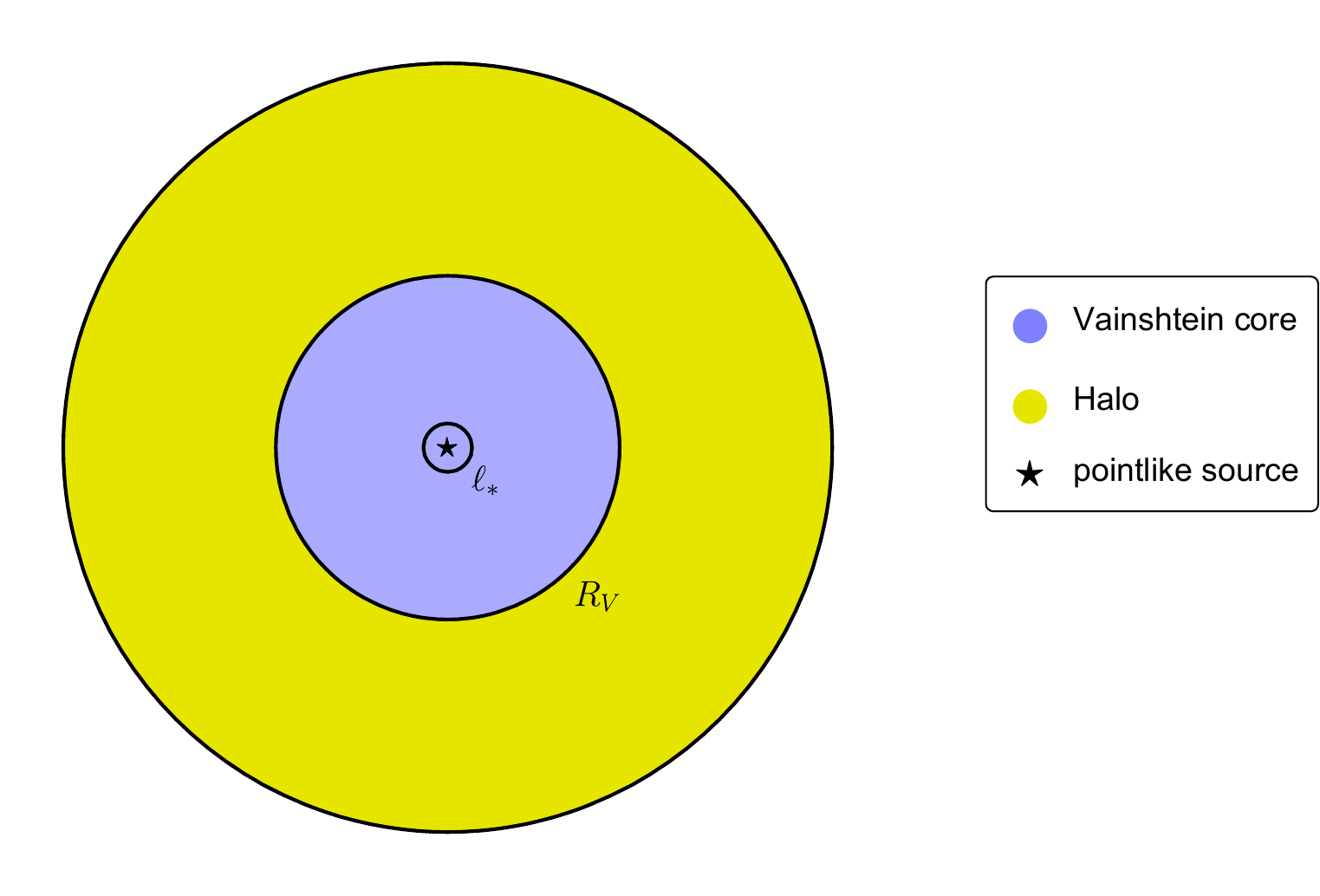}
    \caption{Illustration of the internal structure of a classicalon: a Vainshtein core ($\overline{\rho}^\prime > \Lambda_\ast^2$) surrounded by a halo ($\overline{\rho} > \Lambda_\ast$). $R_V$ denotes the Vainshtein radius, and $\ell_\ast \equiv 1/\Lambda_\ast$ the length scale governing the derivative self-interaction.}
    \label{fig:halo}
\end{figure}

We seek a way to cure the aforementioned strong-coupling problem between $\phi$ and $A_\mu$ without modifying the $P(X)$ structure of the kinetic and derivative self-interaction terms of the Higgs field in Eq.~\eqref{eq:PX_Lag_0} with $n=2$. It is instructive to note that the EFT breakdown actually occurs for $r \gg R_V$. Indeed, $|\phi|=\rho/\sqrt{2} \gg \Lambda_\ast$ in a region---the ``halo''---larger than that delimited by the Vainshtein radius---the ``Vainshtein core''---which begins at
\begin{equation}
r = \ell_\ast \left( \frac{\mathcal{E}}{\Omega\Lambda_\ast} \right) \gg R_V, \ \text{since} \ \mathcal{E} \gg \Lambda_\ast,
\end{equation}
see Fig.~\ref{fig:halo}, where the effective 4-photon interaction in Eq.~\eqref{eq:Lint 4},
\begin{equation}
-\frac{g^4}{4} \underbrace{\left[\frac{\overline{\rho}(r_0)}{\Lambda_{\ast}}\right]^4}_{\gg 1} A^4,
\end{equation}
remains strongly coupled. Hence, a modification of the Lagrangian density \eqref{eq:PX_Lag_0} that cures the EFT breakdown must involve $\phi$, though not necessarily $\partial_\mu \phi$, beyond the terms already present in Eq.~\eqref{eq:PX_Lag_0}.

To cure the strong-coupling problem, we therefore wish to exploit nonlinearities in $\phi$---i.e., a chameleon screening mechanism---similar to that introduced for the Yukawa coupling to fermions in Ref.~\cite{Nortier:2025gmc}, in order to realize a symbiosis with Vainshtein screening \cite{Nortier:2025gmc,Sirera:2026kzx}. To implement this in a gauge-invariant way, we couple the Higgs field to the photon with a gauge kinetic function:
\begin{equation}
    \lagr_\mathrm{cham.} \equiv -\frac{1}{4} e^{a(\zeta)} F^{\mu\nu}F_{\mu\nu},
    \quad \text{with} \quad
    \zeta \equiv \frac{\phi^*\phi}{\Lambda_A^2},
    \label{eq:Lgauge_0}
\end{equation}
where $a(\zeta)$ is an entire function of $\zeta$ and $\Lambda_A$ is a new interaction scale.
To understand why this could cure the strong-coupling problem, we rename the original coupling $g$ as $g_0$ and perform a field redefinition of $A_\mu$:
\begin{equation}
A_\mu \mapsto \frac{A_\mu}{g_0}
\quad \Longrightarrow \quad
D_\mu = \partial_\mu + iA_\mu,
\end{equation}
such that Eq.~\eqref{eq:Lgauge_0} is transformed as
\begin{equation}
    \lagr_\mathrm{cham.} = -\frac{1}{4 g^2(\phi)} F^{\mu\nu}F_{\mu\nu}, \quad
    g(\phi) \equiv e^{-\frac{a(\zeta)}{2}} g_0,
    \label{eq:coupling_rescaling}
\end{equation}
where $g(\phi)$ is an effective gauge coupling depending on $|\phi|$. A scalar background thus acts as a medium with impedance $g^2(\phi)/g_0^2$ for photons. The expectation is that this solves the strong-coupling issue by decoupling the photon from the background when $|\phi| \gg \Lambda_\ast$, by choosing $a(\zeta)$ to be an increasing function of $\zeta \in \mathbb{R_+}$.

We now move to a frame with a canonical photon kinetic term via the field redefinition:
\begin{equation}
A_\mu \mapsto g(\phi) A_\mu,
\label{eq:A_redef}
\end{equation}
such that Eq.~\eqref{eq:coupling_rescaling} is transformed as
\begin{equation}
\lagr_\mathrm{cham.} = - \frac{1}{4} F^{\mu\nu} F_{\mu\nu} + \half \partial^{[\mu} a(\zeta) A^{\nu]} F_{\mu\nu} - \frac{1}{4} \partial^{[\mu} a(\zeta) A^{\nu]} \partial_{[\mu} a(\zeta) A_{\nu]},
\label{eq:Lgauge}
\end{equation}
and Eqs.~\eqref{eq:LX order 1}-\eqref{eq:LX order 4} are modified by the replacement $g \mapsto g(\phi)$.
In what follows, we take $a(\zeta)=\zeta$, since the results generalize to any increasing entire function, and set $\Lambda_A \sim \Lambda_\ast$. Going to unitary gauge in the polar parameterization of the Higgs field \eqref{eq:polar param}, we obtain
\begin{equation}
     \lagr_\mathrm{cham.} = - \frac{1}{4} {F}^{\mu\nu}  {F}_{\mu\nu} + \frac{1}{2\Lambda_{A}^{2}} \rho \partial^\mu \rho A^\nu  F_{\mu\nu} - \frac{1}{8 \Lambda_{A}^{4}} \rho^{2} (\partial \rho)^2 A^2 + \frac{1}{8 \Lambda_{A}^{4}} \rho^2 (\partial^\mu \rho A_\mu)^2.
     \label{eq:Lgauge_2}
\end{equation}
The main difference from the discussion with a Yukawa coupling to a fermion \cite{Nortier:2025gmc} lies in the extra terms generated by the gauge kinetic function \eqref{eq:coupling_rescaling} after the photon field redefinition \eqref{eq:A_redef}. It is therefore important to check whether the scalar fluctuations are well-behaved inside the classicalon.

In Appendix~\ref{fluct_int_terms_cham}, we derive the Lagrangian density for the scalar fluctuations in $\lagr_\mathrm{cham.}$ about the background \eqref{eq:background_real}, assuming a moderately weak coupling $g_0 \lesssim 1$. From the Lagrangian densities \eqref{eq:Lint 2 non lin regime} and \eqref{eq:cham_fluct_Z}---with the replacement $g \mapsto g(\phi)$---inside the Vainshtein core ($\ell_\ast \ll r_0 \ll R_V$), we extract the terms quadratic in $A_\mu$:
\begin{align}
\mathcal{L}_{A^2} \simeq
&- \frac{1}{4} F^{\mu\nu} F_{\mu\nu} - \left( \frac{3r_0}{\ell_\ast} \right) \frac{\sqrt{Z} \Lambda_B}{2} A^\nu F_{r\nu} \nonumber \\
&+ \underbrace{\left( \frac{3r_0}{\ell_\ast} \right)^2 \frac{Z \Lambda_B^2}{8}}_{\gg \Lambda_B^2} \ \ \cdot \underbrace{\left(A^2 + A_r^2\right)}_{\quad = A_t^2-A_\vartheta^2-A_\varphi^2}
+ \ \underbrace{\frac{e^{-Z\left( \frac{3r_0}{\ell_\ast} \right)^2}}{2} \ (M_A^\mathrm{eff})^2}_{\ll \Lambda_B^2} \ \cdot \ A^2,
\end{align}
where the first and second lines give the kinetic and mass terms, respectively. Note that the Lorentz-symmetry breaking by the background is manifest. We observe that, whereas the mass term from the covariant derivative of the Higgs field (last term) is now suppressed, the new mass-like term---arising from the spatial variation of the impedance---dominates for the non-radial components of the photon.

As usual in EFTs, we wish to determine which operator dominates the dynamics in order to organize the power counting of the EFT \cite{Gripaios:2015qya} about the classicalon background. Consider a photon configuration of energy $E$; there are 2 candidates for the kinetic term---a Lorentz-invariant and a Lorentz-breaking one---whose ratio is
\begin{equation}
\left|
\frac{- \left( \frac{3r_0}{\ell_\ast} \right) \frac{\sqrt{Z} \Lambda_B}{2} A^\nu F_{r\nu}}{- \frac{1}{4} F^{\mu\nu} F_{\mu\nu}}
\right|
\sim \left( \frac{r_0}{\ell_\ast} \right) \sqrt{Z} \left( \frac{\Lambda_B}{E} \right) \gg 1, \ \text{for} \ E \lesssim \Lambda_B,
\end{equation}
so we conclude that the Lorentz-breaking term dominates the usual kinetic term below the blueshifted strong-coupling scale $\Lambda_B$. However, when the former is compared with the dominant mass term:
\begin{equation}
\left|
\frac{\left( \frac{3r_0}{\ell_\ast} \right)^2 \frac{Z \Lambda_B^2}{8} (A^2+A_r^2)}{- \left( \frac{3r_0}{\ell_\ast} \right) \frac{\sqrt{Z} \Lambda_B}{2} A^\nu F_{r\nu}}
\right|
\sim \left( \frac{r_0}{\ell_\ast} \right) \sqrt{Z} \left( \frac{\Lambda_B}{E} \right) \gg 1, \ \text{for} \ E \lesssim \Lambda_B,
\end{equation}
we observe that the new mass-like term dominates the leading kinetic term at any energy $E \lesssim \Lambda_B$. The non-radial photon components are pushed above the blueshifted strong-coupling scale of the EFT owing to the large impedance contribution to the Lorentz-breaking effective photon mass. Hence, the symbiosis between Vainshtein and chameleon screening cannot restore the validity of the EFT inside the classicalon, unlike in the Yukawa-coupling case \cite{Nortier:2025gmc}; this symbiosis mechanism therefore cannot be used with a gauge kinetic function in a $P(X)$ Higgs model.

\section{Discussion and Outlook}
\label{generalization}
In this article, we have applied the methodology of Ref.~\cite{Nortier:2025gmc}---summarized in Section~\ref{k-ess_prot} for a simple $P(X)$ k-essence model---to diagnose whether a theory can develop Vainshtein screening, a criterion for self-UV completion via classicalization. We applied this diagnostic to a $P(X)$ generalization of the original Abelian Higgs model, reviewed in Section~\ref{Ab_H_Model}.

In Section~\ref{decoupling_limit}, we showed that, in the limit of vanishing gauge coupling, there is no obstruction to Vainshtein screening around a sharply localized source for the complex scalar field.

In Section~\ref{strong_break_V}, we found that, owing to the coupling between the photon and the Higgs background and its fluctuations, the nonlinearities trigger a strong-coupling breakdown of Vainshtein screening.

In Section~\ref{no_rescue}, we attempted to cure this issue through a symbiosis of chameleon and Vainshtein screening, implemented by coupling the Higgs field to the photon via a gauge kinetic function. However, the new interactions arising from this field-dependent coupling reintroduce a strong-coupling problem, invalidating the attempt.

Below, we briefly discuss how our results may generalize, or be circumvented, in variants of the models studied in detail in this article.

\paragraph{\boldmath General $P(X)$ Higgs Models:}
We focused our analysis on the $P(X)$ model \eqref{eq:PX_Lag_0} with $n=2$. One may wonder whether taking $n>2$ or a more general polynomial $P(X)$ would significantly alter our conclusions. However, the fact that $\rho' \gg \Lambda_\ast^2$ and $\rho \gg \Lambda_\ast$ inside the Vainshtein core is a general feature of $P(X)$ models \cite{deRham:2014wfa,Joyce:2014kja}; hence, the coupling between the scalar background and the photon via covariant derivatives in the kinetic self-interaction term(s) will inevitably lead to the same strong-coupling problem. Moreover, our criticism of the attempt to cure this issue by implementing a symbiosis between the chameleon and Vainshtein mechanisms was independent of the details of the $P(X)$ function. We can therefore confidently extend our conclusions to the class of $P(X)$ Higgs models, including those with a gauge kinetic function.

\paragraph{\boldmath Non-Abelian $P(X)$ Higgs Models:}
Our study focused on an Abelian Higgs model, but since the problems originate from the coupling of the gauge field to the Higgs background, which runs out of control inside the halo and the Vainshtein core, a non-Abelian Higgs model \cite{Kibble:1967sv} will clearly face the same difficulties. Moreover, the criticism of the model in Appendix~\ref{fate_1st_model}, based on the field redefinition in the Abelian example, generalizes straightforwardly to the non-Abelian case. We conclude that generalizing the models explored in this article to realize Higgs classicalization within the EW theory \cite{Weinberg:1967tq,Salam:1968rm} will likewise fail.

\paragraph{Galileon Higgs Models:}
Galileon scalar theories \cite{Luty:2003vm,Nicolis:2004qq,Nicolis:2008in,Joyce:2014kja} realize Vainshtein screening and are therefore natural candidates for classicalizing theories \cite{Dvali:2010jz}. As an alternative to a $P(X)$ Higgs, one may consider a Galileon Higgs \cite{Hull:2014bga}. However, the derivative self-interaction terms expected to trigger Vainshtein screening around a sharply localized source involve covariant derivatives with the gauge fields. Given the standard Galileon backgrounds that realize Vainshtein screening \cite{Nicolis:2004qq,Nicolis:2008in,Joyce:2014kja}, we anticipate the same strong-coupling issues as those encountered for a $P(X)$ Higgs. Nevertheless, a detailed study of the Galileon Higgs would be worthwhile.

\paragraph{Generalized K-chameleon Higgs:}
Although our results disqualify the 2 proposals in the literature for a Higgs classicalizer\footnote{To be precise, there remains the proposal of Ref.~\cite{Chattopadhyay:2023nbj}, based on the covariant star-product formalism of Ref.~\cite{Nortier:2023dkq}, in which Higgs classicalization was argued to occur via infinite-derivative QFT and the classical action is nonlocal. However, it is unclear to us at present what the equivalent of the Vainshtein screening diagnostic would be in that case.} \cite{Dvali:2010jz,Dvali:2016ovn}, this does not amount to a no-go theorem for Higgs classicalization. Indeed, one could construct models with other gauge-covariant operators involving the Higgs field and its derivatives that might avoid the pathologies discussed in this article. For instance, one may consider a more general gauged k-chameleon field in which the kinetic function of the $P(X)$ Lagrangian \eqref{eq:PX_Lag_0} is replaced as follows:
\begin{equation}
P(X) \mapsto \mathcal{K}(\phi^\ast\phi, X),
\end{equation}
where a suitably chosen $\mathcal{K}(\phi^\ast\phi, X)$ might realize a symbiosis between the chameleon and Vainshtein mechanisms without encountering a strong-coupling problem. This is certainly an interesting avenue to explore, even if it ultimately leads to the same pathologies as the $P(X)$ Higgs model. However, an analytical treatment of screening is usually more difficult for a k-essence scalar with a general kinetic function \cite{Brax:2012jr}. Note that this may be 1 possibility among others for circumventing the strong-coupling problem. Establishing a general no-go theorem for Higgs classicalization would definitively rule out this intriguing route to explaining the little hierarchy problem.

\acknowledgments
The authors thank L.~Darmé, A.~Deandrea, M.~Ettengruber, and F.~N.~Mahmoudi for helpful discussions. We acknowledge the use of \texttt{Wolfram Mathematica} for producing Fig.~\ref{fig:halo} and verifying certain formal derivations, and of \texttt{Mistral AI} for assistance with English editing during drafting.

\appendix

\section{Fate of the First Higgs Classicalizer}
\label{fate_1st_model}
The first Higgs-classicalizer model was proposed in Ref.~\cite{Dvali:2010jz}, where the authors suggested triggering classicalization through the operator
\begin{equation}
\dfrac{H^\dagger H}{\Lambda_\ast^2} T_\mu^\mu,
\end{equation}
where $H$ is the EW Higgs doublet and $T_\mu^\mu$ is the trace of the energy-momentum tensor of the SM.

To construct an analogous model for the Abelian Higgs case (Section~\ref{Ab_H_Model}), one may restrict attention to a Lagrangian density containing the minimal terms,
\begin{equation}
\mathcal{L}_\mathrm{class1} \equiv - \frac{1}{4} F^{\mu\nu}F_{\mu\nu} + \left( 1 + \kappa_2 \frac{\phi^*\phi}{\Lambda_\ast^2} \right) (D^\mu \phi)^\ast (D_\mu \phi) - V(\phi),
\label{eq:first_H_class}
\end{equation}
with $\kappa_2 = \pm 1$.
In unitary gauge with the polar parameterization \eqref{eq:polar param}, $\phi(x)=\rho(x)/\sqrt{2}>0$. For a weakly coupled photon, one may test whether the nonlinearities of the scalar derivative self-interaction can trigger classicalization in the scalar sector, described by the Lagrangian density
\begin{equation}
\mathcal{L}_\mathrm{class1}^{\mathrm{scal.}} \equiv \frac{1}{2} \left( 1 + \kappa_2 \frac{\rho^2}{2\Lambda_\ast^2} \right) (\partial \rho)^2 - V(\rho).
\end{equation}

Note that classicalization occurs when the derivative self-interaction term dominates the kinetic one, i.e., when $\rho \gg \Lambda_\ast$. It is then clear that $\kappa_2 = -1$ (as considered in Ref.~\cite{Dvali:2010jz}) would produce a ghost instability for the fluctuations of the background field. In what follows, we therefore focus on $\kappa_2=+1$.

The approach of the authors of Ref.~\cite{Dvali:2010jz} was to study the response of the Higgs field to a sharply localized source, as we do in Section~\ref{PX_Abelian_H} for the $P(X)$ model. One must bear in mind, however, that this external source is a proxy for the backreaction of the derivative self-interaction induced by the localization of hard quanta in a scattering process. One should therefore first verify that this self-sourcing term can indeed trigger classicalization in a dynamical process, such as an incoming spherical wave \cite{Dvali:2010ns,Dvali:2011th}, in which no such artificial probe source is used. For the model defined in Eq.~\eqref{eq:first_H_class}, however, a simple field redefinition ($\rho \mapsto \widetilde{\rho}$),
\begin{equation}
\dfrac{d\widetilde{\rho}}{d\rho} \equiv \sqrt{1 + \frac{\rho^2}{2\Lambda_\ast^2}}, \quad
V(\rho) \mapsto \widetilde{V} \left( \widetilde{\rho} \right),
\end{equation}
reveals that the apparent derivative self-interaction is an artifact of the choice of field-space coordinates:
\begin{equation}
\mathcal{L}_\mathrm{class1}^{\mathrm{scal.}} = \frac{1}{2} \left( \partial \widetilde{\rho} \right)^2 - \widetilde{V} \left( \widetilde{\rho} \right).
\end{equation}
In conclusion, contrary to what a naive analysis of the coupling to an external source might suggest \cite{Dvali:2010jz}, the model \eqref{eq:first_H_class} cannot classicalize the Higgs field. This statement generalizes to any self-interaction operator of the form:
\begin{equation}
- \frac{1}{4} F^{\mu\nu}F_{\mu\nu} + \mathcal{F} \left( \frac{\phi^*\phi}{\Lambda_\ast^2} \right) (D^\mu \phi)^\ast (D_\mu \phi) - V(\phi),
\end{equation}
where $\mathcal{F}(z)$ is an analytic function for $z>0$.

\section{\boldmath Lagrangian Terms of the $P(X)$ Higgs Model}
\subsection{\boldmath Expanded $P(X)$ Higgs Lagrangian}
\label{Lag_exp}
The Lagrangian density $\mathcal{L}_X$ in Eq.~\eqref{eq:PX_Lag_0} (with $n=2$ and $c_2 = -1$) expands in powers of the coupling $g$, hereafter referred to as the ``order'':
\begin{itemize}[label=$\spadesuit$]
    \item order $g^0$:
    \begin{equation}\label{eq:LX order 0}
        \lagr^{(0)}_X \equiv -\frac{1}{4} F^{\mu\nu} F_{\mu\nu} + \pmuu \phi^* \pmud \phi - \frac{1}{\Lambda_\ast^4} \left( \pmuu \phi^* \pmud \phi \right)^2 - V(\phi) + \frac{\phi^\ast}{\Lambda_\ast} T + \frac{\phi}{\Lambda_\ast} T^\ast;
    \end{equation}
    \item order $g^1$:
    \begin{equation}\label{eq:LX order 1}
        \lagr^{(1)}_X \equiv g j^\mu A_\mu - \frac{2g}{\Lambda_\ast^4} \partial^\nu \phi^* \partial_\nu \phi j^\mu A_\mu;
    \end{equation}
    \item order $g^2$:
    \begin{equation}\label{eq:LX order 2}
        \lagr^{(2)}_X \equiv g^2 \phi^*\phi A^2
        - \frac{g^2}{\Lambda_\ast^4} \left( j^\mu j^\nu A_\mu A_\nu + 2 \phi^*\phi \partial^\mu \phi^* \partial_\mu \phi A^2 \right);
    \end{equation}
    \item order $g^3$:
    \begin{equation}\label{eq:LX order 3}
        \lagr^{(3)}_X \equiv -\frac{2 g^3 }{\Lambda_\ast^4} \phi^*\phi j^\mu A_\mu A^2;
    \end{equation}
    \item order $g^4$:
    \begin{equation}\label{eq:LX order 4}
        \lagr^{(4)}_X \equiv -\frac{g^4 }{\Lambda_\ast^4} (\phi^*\phi)^2 A^4,
    \end{equation}
\end{itemize}
where the scalar current $j^\mu$ is defined in Eq.~\eqref{eq:scalar_current}.

\subsection{\boldmath $P(X)$ K-essence Fluctuations Coupled to the Photon}
\label{fluct_int_terms}
From Appendix~\ref{Lag_exp} and using the polar parameterization \eqref{eq:polar param} of the Higgs field, the interaction terms in unitary gauge for $\delta \rho (x) \equiv \rho(x)-\overline{\rho}(r)$, at nontrivial order in $g$, read:
\begin{itemize}[label=$\spadesuit$]
\item Order $g^2$:
\begin{equation}
                \begin{aligned}\label{eq:Lint 2}
                    \lagr_X^{(2)} &= g^2 \left\{ \left[ 1 + \frac{(\overline{\rho}')^2}{\Lambda_\ast^4} \right\} \left( \overline{\rho} \delta\rho + \frac{\delta \rho^2}{2} \right) + \frac{\overline{\rho}^2 \overline{\rho}'}{\Lambda_\ast^4} (\partial_r \delta\rho) - \frac{\overline{\rho}}{\Lambda_\ast^4} \delta\rho (\partial\delta\rho)^2 \right. \\
                    &\quad + \frac{\overline{\rho}'}{\Lambda_\ast^4} (\delta\rho)^2 (\partial_r \delta\rho) - \left. \frac{\overline{\rho}^2}{2\Lambda_\ast^4} (\partial\delta\rho)^2 + \frac{2\overline{\rho}\overline{\rho}'}{\Lambda_\ast^4} \delta\rho (\partial_r \delta\rho) - \frac{1}{2\Lambda_\ast^4} (\delta\rho)^2 (\partial\delta\rho)^2 \right] A^2 \\
                    &\quad + \half \underbrace{ g^2 \overline{\rho}^2 \left[ 1 + \frac{(\overline{\rho}')^2}{\Lambda_\ast^4} \right]}_{\equiv (M_A^{\text{eff}})^2} A^2.
            \end{aligned}
        \end{equation}
We then zoom in on a point $\mathbf{r}=\mathbf{r}_0$ at resolution $\Delta r \ll r_0$, neglecting corrections of order $\Delta r/r_0$ and higher.
In the nonlinear regime ($r_0 \ll R_V$), we perform the field-strength renormalization of the scalar fluctuations. From Eqs.~\eqref{eq:background_real}, \eqref{eq:blueshift} and \eqref{eq:Z_factor}, we obtain $\overline{\rho}(r_0) \simeq \left(\frac{3r_0}{\ell_\ast}\right) \Lambda_B(r_0)$; hence
        \begin{equation}
            \begin{aligned}\label{eq:Lint 2 non lin regime}
                 \lagr_X^{(2)} \simeq g^2 \Bigg[& \left( \frac{3r_0}{\ell_\ast} \right) Z \Lambda_\ast \delta\rho_Z +  \left( \frac{3r_0}{\ell_\ast} \right)^2 Z (\partial_r \delta\rho_Z) + \dfrac{\delta\rho_Z^2}{2} -  \left( \frac{3r_0}{\ell_\ast} \right)^2 \frac{(\partial \delta \rho_Z)^2}{2 \Lambda_\ast^2} \\
                 &+ \left( \frac{3r_0}{\ell_\ast} \right)  \frac{2 (\delta \rho_Z) (\partial_r \delta \rho_Z)}{\Lambda_\ast}  - \left( \frac{3r_0}{\ell_\ast} \right) \frac{\delta \rho_Z (\partial \delta \rho_Z)^2}{Z \Lambda_\ast^3} + \frac{(\delta\rho_Z)^2 (\partial_r \delta \rho_Z) }{\Lambda_B^2} \\
                 &- \frac{(\delta \rho_Z)^2 (\partial \delta\rho_Z)^2}{2 \Lambda_B^4}
                 \Bigg] A^2 + \frac{(M^{\mathrm{eff}}_A)^2}{2} A^2
                 + \mathcal{O} \left( \frac{\Delta r}{r_0} \right),
            \end{aligned}
        \end{equation}
with the effective photon mass:
        \begin{equation}\label{eq:m eff non lin regime}
            M_A^{\mathrm{eff}}(r_0) \simeq g \left( \frac{3r_0}{\ell_\ast} \right) Z(r_0) \Lambda_\ast.
        \end{equation}
\item Order $g^4$:
        \begin{equation}\label{eq:Lint 4}
            \begin{aligned}
                \lagr_X^{(4)} =& -\frac{g^4}{4\Lambda_{\ast}^4} \left( \overline{\rho}^4 + 4\overline{\rho}^3 \delta\rho + 6\overline{\rho}^2 (\delta\rho)^2 + 4\overline{\rho} (\delta\rho)^3 + \delta\rho^4 \right)  A^4.
            \end{aligned}
        \end{equation}
In the nonlinear regime ($r_0 \ll R_V$), following the same steps as for order $g^2$, we obtain
        \begin{equation}
        \begin{aligned}\label{eq:Lint 4 non lin regime}
            \lagr_{X}^{(4)} \simeq -g^4 \Bigg[ &\left( \frac{3r_0}{\ell_\ast} \right)^4 \frac{Z^2}{4} +  \left( \frac{3r_0}{\ell_\ast} \right)^3 Z \frac{\delta\rho_Z}{\Lambda_\ast} + \frac{3}{2} \left( \frac{3r_0}{\ell_\ast} \right)^2 \frac{\delta\rho_Z^2}{\Lambda_\ast^2} \\
            &+ \left( \frac{3r_0}{\ell_\ast} \right) \frac{\delta\rho_Z^3}{Z \Lambda_\ast^3} + \frac{\delta\rho_Z^4}{\Lambda_B^4}  \Bigg]  A^4
            + \mathcal{O} \left( \frac{\Delta r}{r_0} \right).
        \end{aligned}
    \end{equation}
\end{itemize}

\subsection{\boldmath $P(X)$ K-chameleon Fluctuations Coupled to the Photon}
\label{fluct_int_terms_cham}
From Eq.~\eqref{eq:Lgauge_2}, the Lagrangian density $\mathcal{L}_\mathrm{cham.}$ written in terms of the scalar fluctuations $\delta \rho(x) \equiv \rho(x) - \overline{\rho}(r)$ reads
\begin{equation}
    \begin{aligned}
      \lagr_\text{cham.} = &- \frac{1}{4}  {F}_{\mu\nu}  {F}^{\mu\nu} \\
      &+\frac{ {A}_\nu {F}^{\mu\nu} (\delta\rho) (\pmud \delta\rho)}{2 \Lambda_A^2} + \frac{ {A}_\nu {F}^{\mu\nu} (\pmud \delta\rho) \bar{\rho}}{2 \Lambda_A^2} - \frac{ {A}^\nu {F}_{r\nu} (\delta\rho) \bar{\rho}'}{2 \Lambda_A^2} - \frac{ {A}^\nu {F}_{r\nu} \bar{\rho} \bar{\rho}'}{2 \Lambda_A^2} \\
      &+ \frac{ (\delta\rho)^2 (A^\mu \pmud \delta\rho)^2}{8 \Lambda_A^4} + \frac{ \delta\rho (A^\mu\pmud \delta\rho)^2 \bar{\rho}}{4 \Lambda_A^4} + \frac{ (A^\mu \pmud \delta\rho)^2 \bar{\rho}^2}{8 \Lambda_A^4} \\
      &- \frac{ ({A}^\mu \pmud \delta\rho) {A}_r (\delta\rho)^2 \bar{\rho}'}{4 \Lambda_A^4} - \frac{ ({A}^\mu \pmud \delta\rho) {A}_r (\delta\rho) \bar{\rho} \bar{\rho}'}{2 \Lambda_A^4} - \frac{ ({A}^\mu \pmud \delta\rho) {A}_r \bar{\rho}^2 \bar{\rho}'}{4 \Lambda_A^4} \\
      &+ \frac{ {A}_r^2 (\delta\rho)^2 (\bar{\rho}')^2}{8 \Lambda_A^4} + \frac{ {A}_r^2 (\delta\rho) \bar{\rho} (\bar{\rho}')^2}{4 \Lambda_A^4} + \frac{ {A}_r^2 \bar{\rho}^2 (\bar{\rho}')^2}{8 \Lambda_A^4} \\
      &- \frac{ {A}^2 (\delta\rho)^2 (\partial \delta\rho)^2}{8 \Lambda_A^4} - \frac{ {A}^2 (\delta\rho) (\partial \delta\rho)^2 \bar{\rho}}{4 \Lambda_A^4} - \frac{ {A}^2 (\partial \delta\rho)^2 \bar{\rho}^2}{8 \Lambda_A^4} \\
      &+ \frac{ {A}^2 (\delta\rho)^2 (\partial_r \delta\rho) \bar{\rho}'}{4 \Lambda_A^4} + \frac{ {A}^2  (\delta\rho)   (\partial_r \delta\rho)   \bar{\rho}   \bar{\rho}'}{2 \Lambda_A^4} + \frac{ {A}^2 (\partial_r \delta\rho) \bar{\rho}^2   \bar{\rho}'}{4 \Lambda_A^4} \\
      &+ \frac{ {A}^2   (\delta\rho)^2   (\bar{\rho}')^2}{8 \Lambda_A^4} + \frac{ {A}^2   (\delta\rho)   \bar{\rho}   (\bar{\rho}')^2}{4 \Lambda_A^4} + \frac{ {A}^2   \bar{\rho}^2   (\bar{\rho}')^2}{8 \Lambda_A^4}.
    \end{aligned}
\end{equation}
We then zoom in on a point $\mathbf{r}=\mathbf{r}_0$ at resolution $\Delta r \ll r_0$, neglecting corrections of order $\Delta r/r_0$ and higher.
Inside the Vainshtein core ($\ell_\ast \ll r_0 \ll R_V$), we perform the field-strength renormalization \eqref{eq:blueshift} of the scalar fluctuations about the background \eqref{eq:background_real}. Using $\Lambda_A \sim \Lambda_\ast$ and the asymptotic behavior $\overline{\rho}(r_0) \simeq \sqrt{Z(r_0)}\Lambda_{\ast}\left(\frac{3r_0}{\ell_\ast}\right)$, we obtain
\begin{equation}
    \begin{aligned}
      \lagr_\text{cham.} \simeq &- \frac{1}{4}  {F}_{\mu\nu}  {F}^{\mu\nu}
      +\frac{ {A}_\nu   {F}^{\mu\nu} (\delta\rho_Z) (\pmud \delta\rho_Z)}{2 Z\Lambda_\ast^2}  \\
      &+ \left(\frac{3r_0}{\ell_\ast}\right) \frac{ {A}_\nu {F}^{\mu\nu} (\pmud \delta\rho_Z)}{2\Lambda_\ast} - \frac{ {A}^\nu   {F}_{r\nu}  (\delta\rho_Z) }{2} - \left(\frac{3r_0}{\ell_\ast}\right) \frac{Z\Lambda_{\ast}   {A}^\nu   {F}_{r\nu}}{2}  \\
      &+ \frac{ (\delta\rho_Z)^2   ({A}^\mu \pmud \delta\rho_Z)^2 }{8 Z^2 \Lambda_\ast^4}  + \left(\frac{3r_0}{\ell_\ast}\right) \frac{ \delta\rho_Z   ({A}^\mu \pmud \delta\rho_Z)^2 }{4Z \Lambda_\ast^3} + \left(\frac{3r_0}{\ell_\ast}\right)^2 \frac{ ({A}^\mu \partial_\mu \delta\rho_Z)^2}{8\Lambda_\ast^2} \\
      &-\frac{ {A}_r   (\delta\rho_Z)^2   ({A}^\mu \partial_\mu \delta\rho_Z)}{4 Z \Lambda_\ast^2} - \left(\frac{3r_0}{\ell_\ast}\right) \frac{  {A}_r   (\delta\rho_Z)   ({A}^\mu\partial_\mu \delta\rho_Z)}{2\Lambda_\ast} - \left(\frac{3r_0}{\ell_\ast}\right)^2 \frac{Z   {A}_r   ({A}^\mu \pmud \delta\rho_Z)}{4}  \\
      &+ \frac{ {A}_r^2   (\delta\rho_Z)^2 }{8} + \left(\frac{3r_0}{\ell_\ast}\right) \frac{Z\Lambda_\ast   {A}_r^2   (\delta\rho_Z)}{4}  + \left(\frac{3r_0}{\ell_\ast}\right)^2 \frac{Z^2 \Lambda_\ast^2   {A}_r^2}{8}  \\
      &- \frac{ {A}^2   (\delta\rho_Z)^2 (\partial \delta\rho_Z)^2}{8 Z^2 \Lambda_\ast^4} - \left(\frac{3r_0}{\ell_\ast}\right) \frac{ {A}^2   (\delta\rho_Z) (\partial \delta\rho_Z)^2}{4Z \Lambda_\ast^3} - \left(\frac{3r_0}{\ell_\ast}\right)^2 \frac{ {A} ^2 (\partial \delta\rho_Z)^2 }{8\Lambda_\ast^2} \\
      &+ \frac{ {A}^2   (\delta\rho_Z)^2   (\partial_r \delta\rho_Z)}{4 Z \Lambda_\ast^2} + \left(\frac{3r_0}{\ell_\ast}\right) \frac{ {A}^2   (\delta\rho_Z)   (\partial_r \delta\rho_Z)}{2\Lambda_\ast} + \left(\frac{3r_0}{\ell_\ast}\right)^2 \frac{Z {A}^2   (\partial_r \delta\rho_Z)}{4}    \\
      &+ \frac{ {A}^2   (\delta\rho_Z)^2}{8} + \left(\frac{3r_0}{\ell_\ast}\right)  \frac{Z\Lambda_\ast   {A}^2   (\delta\rho_Z)}{4} + \left(\frac{3r_0}{\ell_\ast}\right)^2 \frac{Z^2 \Lambda_{\ast}^2   {A}^2}{8}
      + \mathcal{O} \left( \frac{\Delta r}{r_0} \right).
    \end{aligned}
\label{eq:cham_fluct_Z}
\end{equation}

\bibliographystyle{JHEP}

\providecommand{\href}[2]{#2}\begingroup\raggedright\endgroup

\end{document}